# Multi-Tier Mentorship with AI-Assisted Development: Authentic Engineering for K-12 and Undergraduates

Kelly Yuan, Ronald Liu, Daniel Crawford, and Weihao Qu*
kellyyuan@ctemc.org, rolandyoyang@icloud.com, s1323702,wqu@monmouth.edu

***Abstract*** **– K-12 students often possess creative engineering ideas but lack technical skills to build them, while undergraduates have coding expertise but few opportunities to lead real-world projects or mentor others. The rapid development of AI-assisted tools offers a potential bridge to connect these groups, yet the structure for effective K-12 and university collaborations remains underexplored. This paper introduces a multi-tiered mentorship framework enabling high school students to engage in authentic engineering through AI-assisted development using large language models and AI agents, while undergraduate mentors provide architectural oversight. We test this framework through LuckyTag, a privacy-preserving NFC-based lost-and-found system. The model positions high schoolers as product leads, undergraduates as technical architects, and faculty as minimal-intervention advisors. A pilot with four high school students, three undergraduates and two faculty yielded survey data showing high perceived barrier removal and gains in system architecture understanding. Thematic analysis reveals that AI amplifies rather than supplants mentoring demands, requiring human oversight for logic and security. These findings suggest a hybrid model for equitable K-12 and university collaboration on computing integration that emphasizes "AI micromanagement" and architectural reasoning over traditional syntax.**

## INTRODUCTION

Pre-college engineering education emphasizes problem decomposition, constrained prototyping, and iterative evaluation [1, 2]. However, persistent syntax and implementation barriers often confine high school students to peripheral roles in software development [3, 4]. Meanwhile, undergraduates computing majors acquire technical proficiency with large language models (LLMs) and AI code generation tools [5] but rarely exercise leadership in mentoring novices or overseeing deployable systems [6]. The advent of accessible generative AI presents an opportunity to bridge these gaps, enabling natural language-driven prototyping that shifts the novice focus from boilerplate coding to higher-order design reasoning [7, 8]. However, these tools risk promoting superficial prompting over robust engineering judgment without structured oversight [9], especially in security-sensitive domains [10].

This work originates from a concrete failure. During a family trip to Switzerland, a high school student on our team, Kelly, lost her iPhone on a regional train near Lauterbrunnen and was unable to recover it despite contacting the rail operator, local police, and tourism office. Apple's "Find My" service proved insufficient once the device went offline. There was no privacy-preserving way for a good-faith finder to reach the owner, and sensitive identifiers (such as a debit card or photo in the phone case) remained hidden where exposing them would create safety risks. This experience led Kelly to articulate the design requirements for a privacy-first lost-and-found service and seek technical collaborators to realize it through AI-assisted development with undergraduate mentors and university advisors.

This paper investigates whether a multi-tiered mentorship model, integrating AI-assisted development with differentiated roles, can enable high school students to perform authentic engineering activities while fostering undergraduate leadership. We test this through LuckyTag, a privacy-preserving NFC lost-and-found system addressing finder liability and owner data exposure. High school participants act as product leads, articulating requirements and generating prototypes via LLMs. Undergraduates serve as technical architects, validating architecture and security. Faculty provide milestone guidance. A one-semester pilot with four high school students (grades 9-12) with limited prior coding experience and three undergraduate Computer Science mentors produced survey data revealing high perceived barrier removal from AI, and elevated post-intervention system architecture understanding, as well as emergent human-AI collaboration challenges. Our contributions include (1) a replicable framework for hybrid AI-assisted education across educational levels (connecting high school and undergraduate students), (2) empirical evidence challenging LLM deskilling narratives, and (3) a research agenda for structured AI pedagogies in privacy-first application contexts.

The remainder of this paper is organized as follows: Section 2 reviews related work. Section 3 details the framework; Section 4 presents the LuckyTag case. Section 5 describes the prompts as engineering artifacts. Sections 6 and 7 report results and discussion, and Section 8 concludes with implications for integrated STEM education.

## Related Work

### I. Mentorship Models in STEM Education

Near-peer mentorship [11] enhances K-12 computing engagement [12] by providing accessible technical guidance, though most programs confine pre-college participants to exploratory tasks rather than full-cycle engineering. Mentorship plays a critical role in the retention, success, and identity development of students within STEM disciplines, particularly for those transitioning from pre-college to higher education environments [13]. Effective STEM education extends beyond curriculum delivery to include social and academic support systems that foster a sense of belonging. Selective STEM high schools frequently support student learning through apprenticeships with scientists, which has been associated with a higher likelihood of students pursuing and completing STEM majors compared to peers without such experiences [14].

### II. AI-Assisted Tools in Computing Education

Generative AI tools such as GitHub Copilot and Large Language Models (LLMs) lower syntactic barriers, enabling novices to prototype complex features via natural language prompts[15]. Research on AI-assisted programming demonstrates that these tools enhance development productivity for experienced practitioners while lowering entry barriers for novices [16]. However, learners must cultivate new competencies such as code evaluation, diagnostic debugging, and architectural decision-making that span human and AI-generated components [17, 18]. K-12 applications focus on isolated exercises rather than integrated hardware-software-ethics challenges [19].

### III. Current Limitations and Research Gap

Project-based learning promotes engineering practices, particularly effective in STEM disciplines when utilized in small groups [20]. Real-world contexts in these projects significantly boost student motivation, yet scaling these efforts to deployable full-stack applications remains a persistent challenge [21]. Because of implementation limitations, novices are often relegated to design-only roles, while some other critical considerations receive minimal attention including security, privacy, and robust engineering judgment.

While individual strands of near-peer mentorship, AI-assisted code generation, and project-based engineering [22] appear separately in the literature, a significant gap remains: there is no integrated, multi-tier framework that enables high school students to lead the development of deployable, security-sensitive systems. Current AI tools facilitate a technical shift from boilerplate coding to design reasoning, but they lack the formal mentorship scaffolding required to translate "vibe coding" into reliable, verified engineering outcomes [23]. Our work explores this direction by coupling AI-assisted development with differentiated mentorship roles, providing empirical insights into how structured human-AI collaboration can navigate traditional implementation barriers in K-12 environments.

## The Multi-Tier Mentorship Framework

The framework developed for this study distributes cognitive load and technical risk through three differentiated roles, creating a structured environment for authentic engineering. This model was approved by the Institutional Review Board (IRB) at Monmouth University to ensure ethical standards in student engagement and data collection.

### I. Recruitment and Context

High school participants were recruited from the Monmouth YOUTH Communication Club, a faculty-supervised initiative for local high school students. Recruitment was conducted through project-specific posters and targeted emails. Undergraduate mentors are selected through faculty invitations. Monmouth University students who demonstrated interest in summer research and engineering projects were invited to participate as technical leads. The study followed a two-phase timeline: **Phase I (Summer 2025)**: An intensive, high-frequency period from June to August characterized by frequent in-person meetings between mentors and high school participants. **Phase II (Fall 2025)**: Implementation and refinement from September to December. Due to academic schedules, collaboration shifted to a remote-first model utilizing Zoom and online collaboration tools.

The framework defines three interdependent roles to distribute cognitive load while preserving student agency as shown in Figure I:

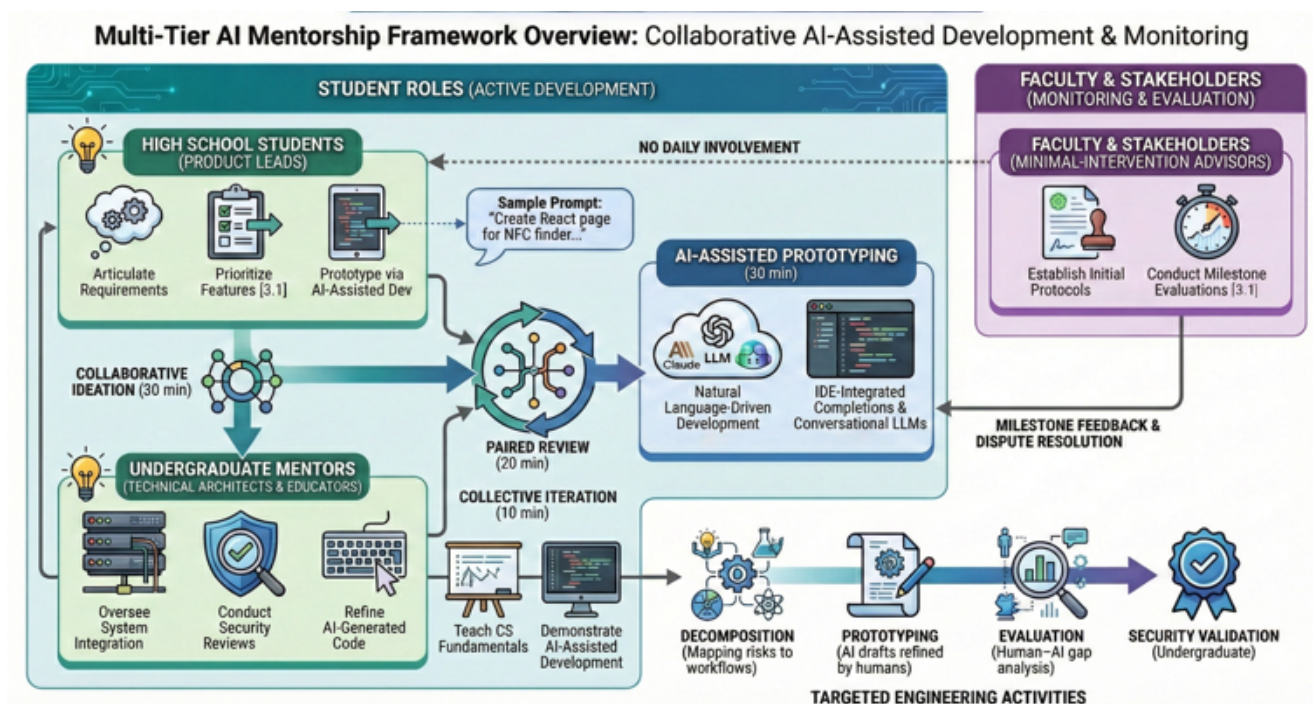

FIGURE I
MULTI-TIER AI MENTORSHIP FRAMEWORK

- **High School Participants (Product Leads):** Students focus on problem articulation, user experience, and initial implementation. Using natural language prompts, they generate functional prototypes and iterate on feature sets. This role emphasizes the "intent" of the system.
- **Undergraduate Mentors (Technical Architects):** Mentors provide the architectural scaffolding. They conduct security audits, validate AI-generated logic, and abstract complex technical roadblocks (e.g., database schema design or API integration) into manageable tasks for the HS students.

- **Faculty (Advisors):** Faculty provide milestone-level governance and strategic oversight. They ensure the project adheres to pedagogical goals and safety standards without interfering in the daily technical collaboration.

Figure I illustrates the information flow between these roles. The HS students initiate the cycle by articulating product requirements. This "intent" is processed through AI-assisted development tools to create a functional baseline. The Undergraduate mentors then review this baseline for architectural integrity and security, providing feedback that loops back to the HS student for further iteration. This creates a continuous cycle of creation and verification, monitored at the perimeter by Faculty advisors.

*II. Pedagogical Scaffolding: The "Froggy-to-AI" Pipeline*

To ensure that high school students could effectively evaluate AI-generated code, undergraduate mentors established a technical baseline before introducing generative tool using steps. This first step is **Fundamental Training**: Mentors introduced core frontend concepts (CSS and layout) through gamified learning platforms such as *Flexbox Froggy* as shown in Figure II. This allowed students to internalize the logic of web design in a low-stakes environment.

In the second **Demonstration and Hand-off** stage, mentors demonstrated specific technical workflows, such as API integration or repository management. The third step is **AI Integration**: Once students possessed the conceptual vocabulary, they began using AI-assisted development tools. This sequence ensured that the AI was used to augment existing knowledge rather than replace it, allowing students to identify when the AI output deviated from the intended layout or logic. Notably, mentorship was not unidirectional. Undergraduate mentors reported learning from the high school students' creativity and modern design intuitions.

.

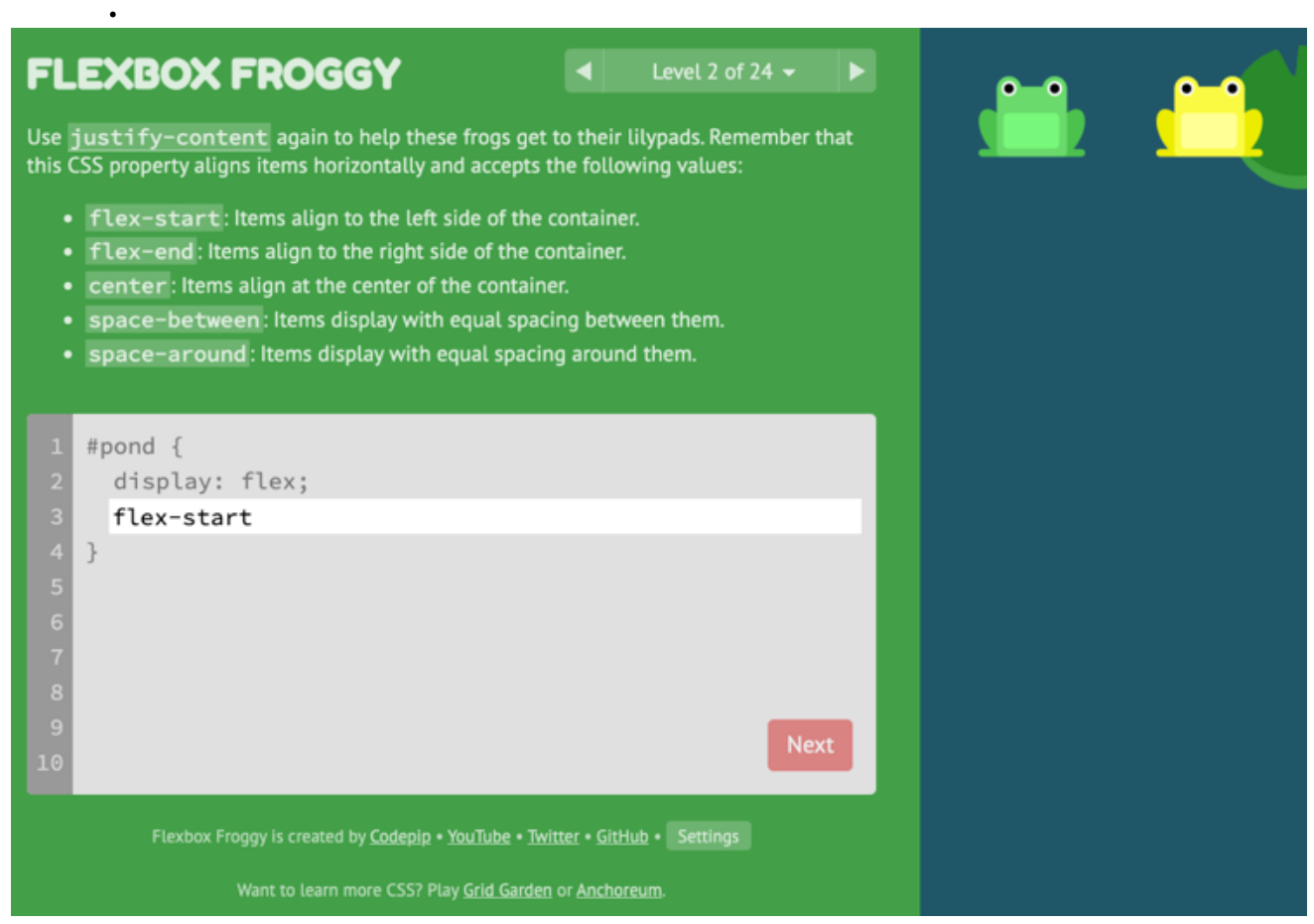

FIGURE II
SCREENSHOT OF FLEXBOX FROGGY GAME

## CASE STUDY: LUCKYTAG

LuckyTag serves as the primary implementation of the multi-tier mentorship framework. The project addresses the real-world problem of lost property recovery through a privacy-preserving, hardware-integrated software system.

*I. Origin: From Lost Phone to Product Vision*

LuckyTag originated from Kelly's experience losing her iPhone on a Swiss train. Systemic failures such as offline tracking, language barriers, closed police stations, and no privacy-safe finder contact, prevented recovery despite diligent efforts. Her initial concept specified QR-tagged items, activation flows, and dialog-based workflows enabling finders to report details (what/where/when) without exposing owner information. This student-led specification framed core requirements: trusted drop-offs, anonymized communication, ownership verification, and prompted seeking AI-assisted implementation support.

The core design philosophy of LuckyTag is "minimal data, maximal trust." Unlike traditional recovery systems that rely on visible phone numbers or addresses, LuckyTag utilizes an anonymized notification pipeline.

- **Owner Privacy**: Personal identifiers are never stored on the physical tag.
- **Finder Safety**: Finders are guided to trusted "Hubs" (e.g., University front desks) rather than meeting owners directly, mitigating liability and safety risks.

*II. Technical Architecture and Stack*

The LuckyTag system employs a professional-grade full-stack architecture designed to balance high-performance deployment with the rapid iteration cycles facilitated by AI-assisted development. For the **frontend**, the team utilized Next.js and React to construct a responsive, mobile-first interface, ensuring that finders in the field can interact with the platform seamlessly across various mobile devices. This client-side infrastructure is integrated with a **backend** powered by Node.js and a Supabase database, which provides the robust foundation necessary for real-time item status updates and secure, anonymized message routing. This specific stack was chosen not only for its scalability but also for its high compatibility with AI code generation tools, which allowed the high school students to prototype complex UI components while undergraduate mentors focused on the underlying logic and data security.

To bridge the physical and digital domains, the system incorporates NFC tags as the primary hardware interface as shown in Figure III. These tags were selected due to their universal compatibility with modern smartphone hardware and their superior physical durability compared to traditional optical identifiers, ensuring the system remains functional in high-wear environments.

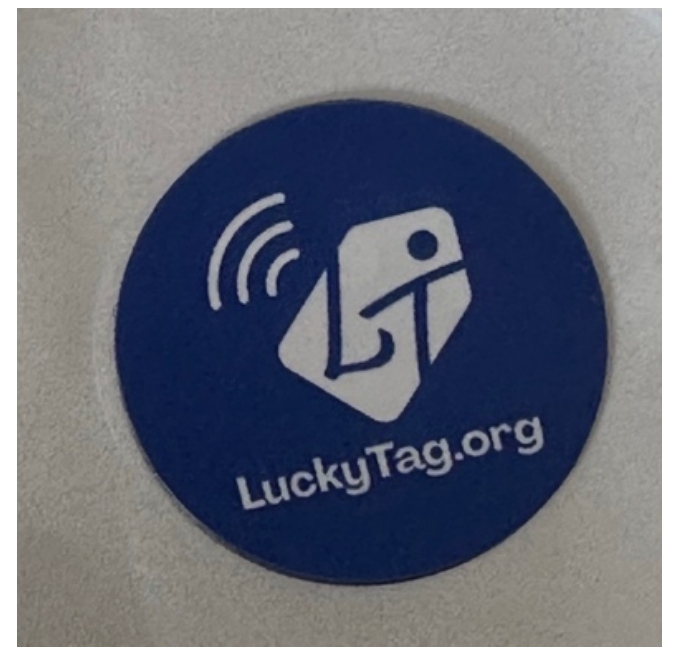

(a) NFC tag

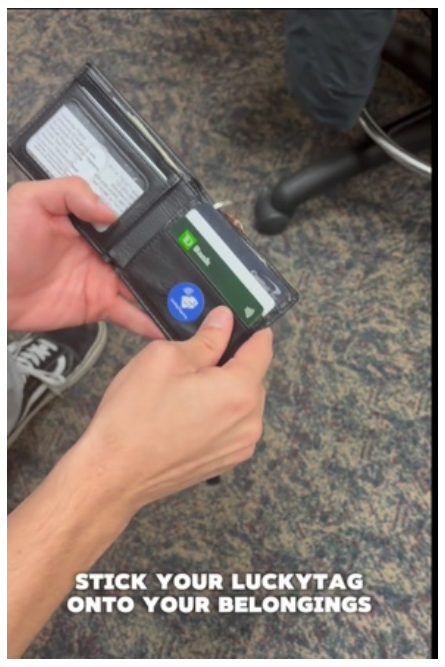

(b) NFC tag usage

FIGURE III
LUCKYTAG NFC TAG

*III. The Engineering Pivot: QR Codes vs. NFC*

A pivotal engineering decision occurred during the transition from the Summer to the Fall semester. The initial prototypes utilized QR codes for item tracking due to the ease of generation within the AI-assisted environment. However, a technical review by the undergraduate mentors identified significant "security-sensitive" flaws as listed in Figure IV such as (1) Vulnerability to "QRjacking": QR codes can be easily covered by malicious stickers, leading users to phishing sites. (2) Physical Durability: Optical codes are prone to scratching and fading on high-wear items like keychains.

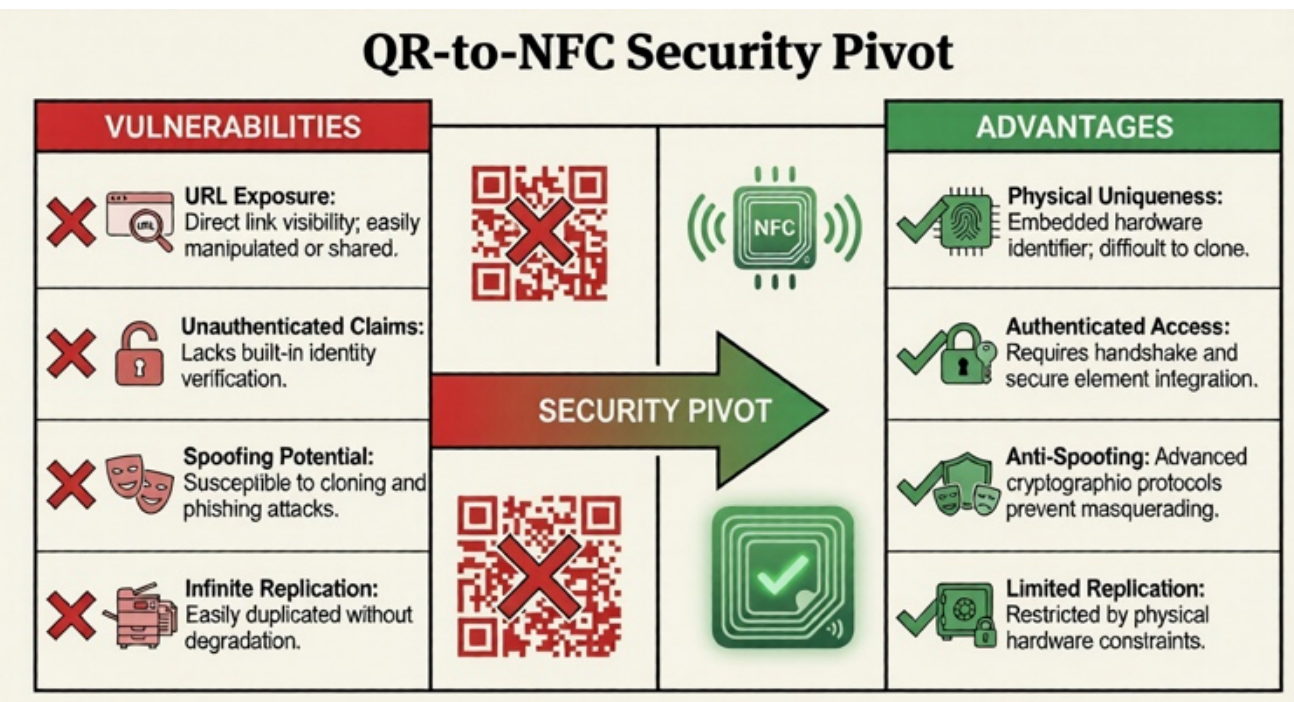

FIGURE IV
QR-NFC COMPARISON

Following this review, the undergraduate mentors provided the architectural scaffolding for an NFC-based system. The high school students then re-engineered the reporting logic to utilize the unique identifier (UID) inherent in each NFC chip. This pivot demonstrated a high-order engineering trade-off: sacrificing "ease of generation" (QR) for "system integrity and security" (NFC).

*IV. Operational Workflow*

The resulting system facilitates a secure four-step recovery process, visualized in Figure V: (1) owners register and link physical tags to accounts; (2) finders scan tags to view nearby trusted drop-offs (libraries, police stations) without owner details; (3) staff authenticate check-ins via portal, triggering owner notifications; (4) owners retrieve items after verification. Role-based access via JWT ensures staff-only sensitive operations.

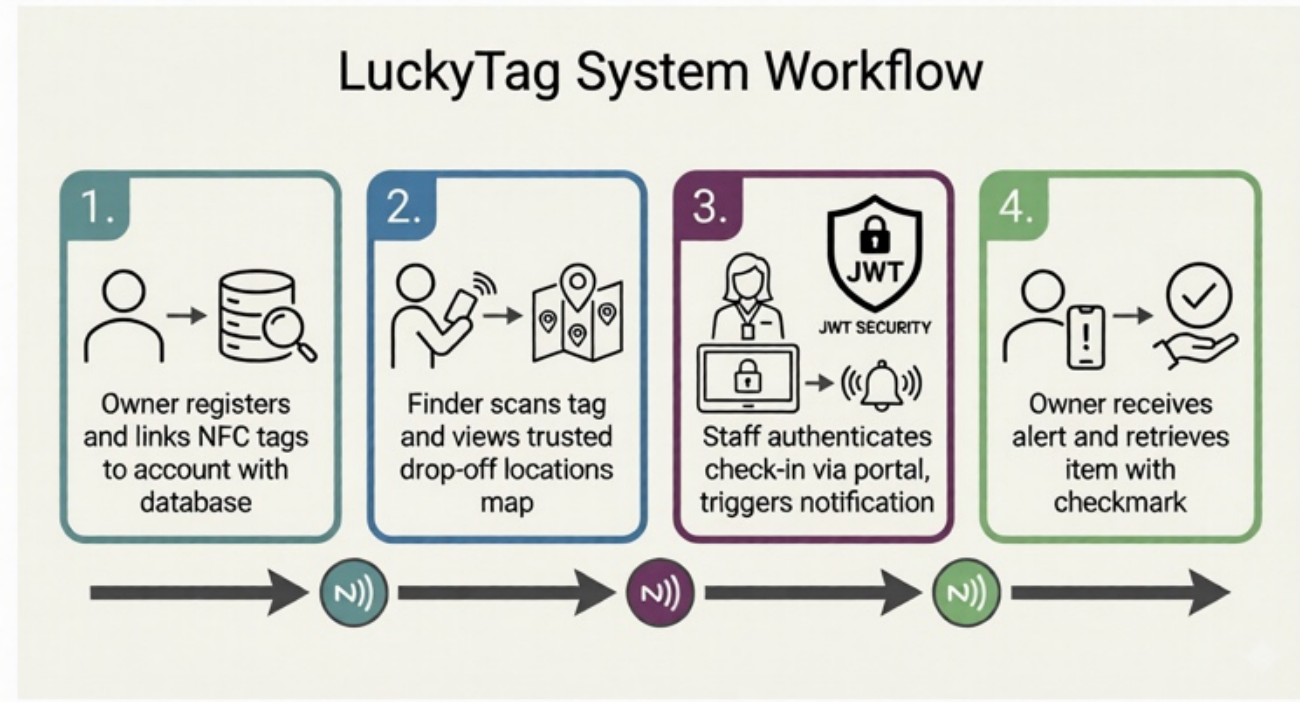

FIGURE V
LUCKYTAG SYSTEM WORKFLOW

## THE PROMPT AS AN ENGINEERING ARTIFACT

To evaluate the depth of student engagement, we analyzed natural language prompts during the development of LuckyTag, as primary engineering artifacts. These documents prove that while AI tools managed the syntax, the high school participants performed the critical roles of requirement articulation, system decomposition, and technical lifecycle management. The following prompts, provided by a Grade 11 participant, illustrate the evolution of the LuckyTag system from initial intent to final deployment.

*I. Initial System Decomposition*

The project began with a comprehensive starting prompt that defined the entire application architecture. Rather than requesting a generic solution, the student articulated a multi-role system with a defined state machine for the tracked items.

***Student Prompt 1 (System Baseline):*** *> "This is a lost and found system. NFC tag is placed on item to keep track of them... The owner page—user can sign up, manage tag status (active, lost, found, picked-up, discarded)... The finder page—user can scan NFC to find the closest locations... The location staff page—user can input location and contact info... scan NFC, be able to notify owner using in-system messaging and also send email."*

This prompt demonstrates **System-Level Thinking**. The student identified three distinct user personas (Owner, Finder, Staff) and a comprehensive state machine for the hardware (Active, Lost, Found, etc.). By being "extra specific," the student acted as a Lead Systems Engineer, ensuring the AI-assisted development tool remained within the bounds of a professional, functional architecture. The main page and the finder paper described in the prompt were constructed by AI agent, as shown by Figure VI(a) and Figure VI(b).

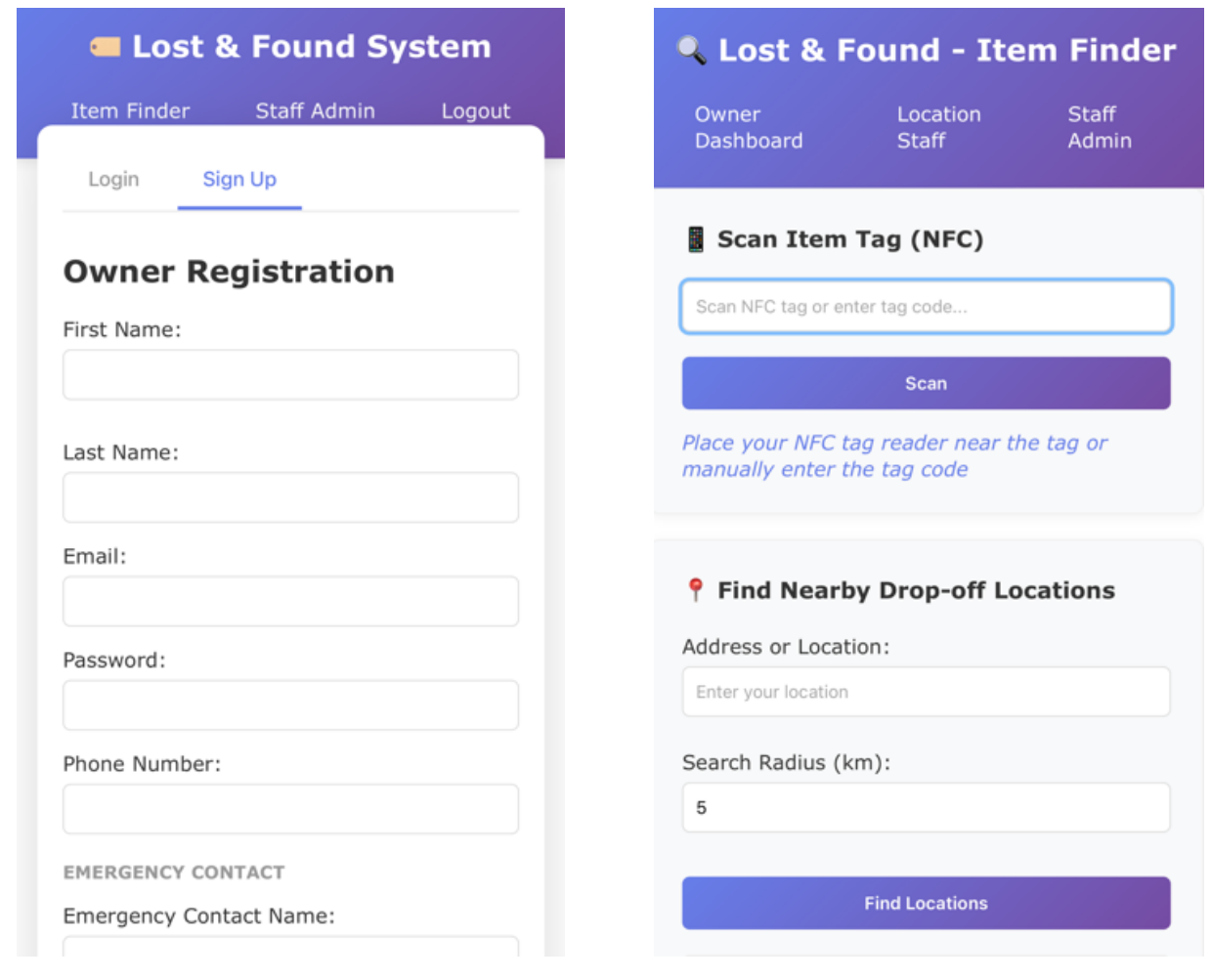

(a) main page (b) finder page

FIGURE VI
SELECTED SCREENSHOT AFTER PROMPT 1

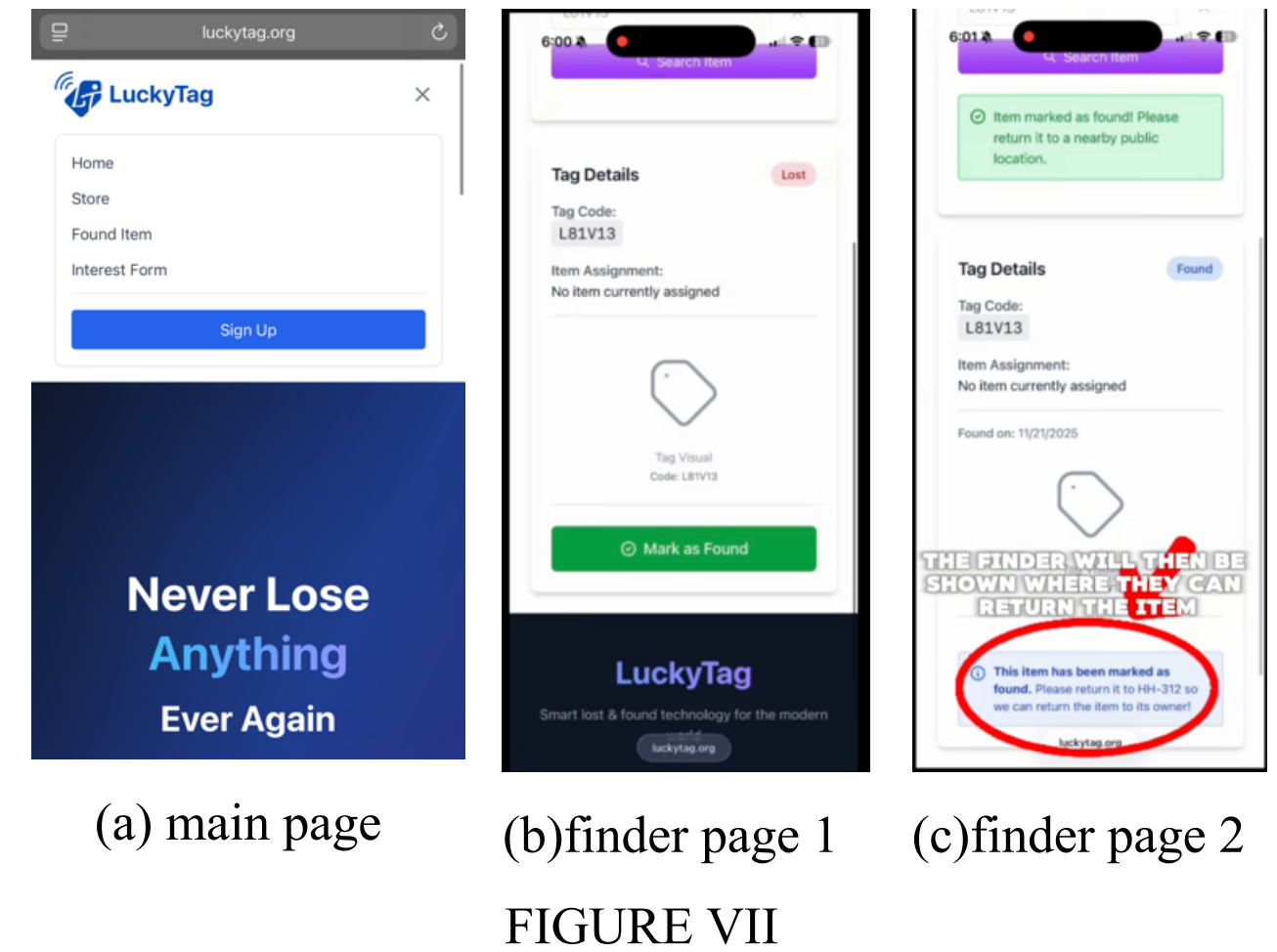

(a) main page (b)finder page 1 (c)finder page 2

FIGURE VII
SELECTED SCREENSHOT AFTER PROMPT 3

*II. Technical Lifecycle and DevOps*

Midway through the project, the focus shifted from feature creation to deployment. The student successfully navigated the transition from a local development environment to an external version control system.

***Student Prompt 2 (Deployment):*** *> "Can you get my app ready to push to GitHub. Once it's ready, push it to this repository: [GitHub URL]."*

Navigating external repositories is typically a significant hurdle for novice coders. This artifact shows the student's growing DevOps Literacy. In a traditional setting, pushing a full-stack app to a remote repository often poses a significant barrier to novices. Here, the student used AI to manage the configuration while maintaining control over the project's lifecycle and versioning.

*III. Iterative Refinement and UX Design*

The final phase involved refining the professional identity of the product. The student took ownership of the branding and user experience (UX) by providing specific assets and naming conventions.

***Student Prompt 3 (Branding):*** *> "Can you change the name of the website to 'LuckyTag' and use this image as the logo for it?"*

This demonstrates Iterative Evaluation. The student moved beyond functional requirements to aesthetic and professional standards, ensuring the final output matched the "privacy-first" identity established in the design phase. After the Branding prompt, the LuckyTag system is updated as shown in Figure VII.

## METHOD

As LuckyTag is in its initial deployment phase, evaluation focuses on pedagogical efficacy rather than long-term product metrics. The goal was to assess whether AI-assisted development enables K-12 students to engage with engineering concepts typically reserved for advanced undergraduates.

*I. Participants and Instruction*

The study involved four high school students (grades 9-12, limited prior coding experience) and three undergraduate mentors from a computer science program. Survey respondents totaled nine (n=9): four high school students, three undergraduate mentor, and two university advisors. Evaluation data was collected through three instruments:

- **Self-Efficacy Survey**: Post-project Likert-scale survey (1–5) measuring confidence in full-stack development, system architecture, and security protocols.
- **Conceptual Fluency Interviews**: Code-walkthrough sessions where students explained AI-generated component logic (e.g., JWT token verification).
- **Prompt Decomposition Logs**: Analysis of natural language prompts to assess problem modularization ability.

*II. Analysis*

Quantitative data reports descriptive statistics (means by role). Qualitative responses underwent thematic coding using a predefined codebook with six constructs: self-efficacy, system thinking, human-AI gap, near-peer scaffolding, purpose/empathy, and engineering identity. Table I presents the codebook used for thematic analysis.

TABLE I

CODE BOOK FOR THEMATIC ANALYSIS

| Code | Definition | Inclusion |
|---|---|---|
| Self-Efficacy | Belief in technical capability | “no longer scared,” “more capable” |
| System Thinking | Focus on architecture and flow | “backend”, “ security” |
| Human-AI gap | Manual correction of AI errors | “fix AI code”, “check AI work” |
| Near-Peer Scaffolding | Learning through mentor guidance. | “jargon removal”, “mentor simplified” |
| Purpose/Empathy | Connection to human needs | “helping others”, “mission” |
| Engineering Identity | Engineer as AI manager | “manager”, “oversight”,“planner” |

## RESULT

### I. Quantitative Findings

The quantitative results, summarized in Table II, indicate high levels of perceived efficacy across all participant groups. High school students reported strong barrier removal (M=4.50), confirming that AI-assisted development enables building features beyond manual coding ability. The logical focus score (M=4.25) indicates attention shifted from syntax to system architecture and user safety. Advisors rated curriculum integration highly (M=4.75), suggesting the model successfully combines hardware, software, and ethics.

TABLE II

QUANTITATIVE PERFORMANCE METRICS (LIKERT SCALE 1–5)

| Metric | Global Mean | HS Mean | UG/Advisor Mean |
|---|---|---|---|
| Professional Identity | 4.56 | 4.50 | 4.60 |
| Efficiency (MVP speed) | 4.60 | 4.50 | 4.75 |
| Purpose-drive Learning | 4.29 | 4.33 | 4.25 |
| Barrier Removal | 4.50 | 4.50 | N/A |
| Logical Focus | 4.11 | 4.25 | 4.00 |
| Near-Peer Accessibility | 4.00 | 4.00 | N/A |
| Curriculum Integration | 4.75 | N/A | 4.75 |

Furthermore, the high scores for Professional Identity (M=4.56) suggest that the multi-tier model successfully fostered a sense of "authentic engineering." Participants did not feel like they were merely completing a classroom exercise; they felt like members of a professional development team.

### II. Qualitative Findings

Thematic analysis revealed four primary shifts in participant perspectives.

*Human-AI Gap:* Students developed critical evaluation skills. Five of nine respondents manually corrected AI code. One high school student reported a UI bug where events reappeared after clicking: "*This error was not fixed by other AIs... eventually, I had to change the code myself.*"

*Engineering Identity*: Participants reframed engineering as a managerial role. One undergraduate mentor stated: "*As an engineer, you still need very deep knowledge of system architecture... AI has become a tool that transforms how these problems are solved.*" A high school student described engineers as people who fix "bugs, small errors, and potential mishaps" in AI-generated frameworks.

*Near-Peer Scaffolding*: Mentors removed technical jargon to facilitate learning. One mentor explained: "*I explained things at the highest level so they would gain understanding without going into depth with complex technical jargon.*"

*Purpose as Driver*: Solving real-world problems increased motivation. One student reflected: "*The motivation to learn switches from 'passing a test' to solving a human need... this sense of purpose makes mastering skills feel less like a chore and more like a mission.*"

A recurring theme in the results was the transition from "producing" to "overseeing." Undergraduate mentors reported that their primary task was not writing code, but Logical Verification. This resulted in a high System Architecture Understanding score (M=4.75 for UG), as the mentors were forced to think about how individual AI-generated components fit into the broader LuckyTag ecosystem. This sets the stage for the emergence of the "AI Micromanager" identity discussed in the following section.

## DISCUSSION

The data from the LuckyTag pilot challenges the prevalent narrative that AI-assisted development "deskills" students. Instead, our findings suggest a fundamental shift in what it means to be an engineer in the age of generative models.

### I. The "Fundamentals-to-AI" Pipeline

A critical observation was that AI-assisted development was most effective when preceded by conceptual scaffolding. Undergraduate mentors did not begin with AI. Instead, they established a technical baseline using gamified platforms like Flexbox Froggy. This "Fundamentals-First" approach allowed high school students to internalize CSS and layout logic in a low-stakes environment.

When students transitioned to using AI tools for the LuckyTag frontend, this baseline served as a "sanity check." Because they understood the underlying logic of Flexbox, they could identify when AI-generated components deviated from their design intent. This suggests that for AI-augmented engineering to be successful, students still require a "syntax-lite" conceptual vocabulary to act as informed evaluators rather than passive consumers of AI output.

### II. The Emergence of the "AI Micromanager"

Both undergraduate mentors and high school students reported a shift from "writing code" to "managing logic". While AI automated the syntax for LuckyTag, it lacked the threat modeling intuition to foresee risks like QRjacking or physical durability flaws. As the HS participants identified bugs that the AI could not resolve, such as UI alignment issues or database routing errors, they were forced to engage in high-level debugging. This process mirrors professional engineering leadership, where the primary value is not in writing the lines of code, but in the logical verification of the system's integrity. By overseeing tasks like JWT-based access control and the NFC security pivot, students moved beyond "vibe coding" toward authentic engineering leadership.

### *III. Reciprocal Mentorship and Creative Synergy*

Counterintuitively, the mentorship was not unidirectional. While undergraduate mentors provided technical scaffolding and architectural guidance, they reported learning significantly from the high school students. High school participants, unburdened by legacy coding constraints, often displayed superior creativity and design intuition.

One UG mentor noted that the HS student's vision for a "cleaner, more modern UI" pushed the mentor to explore new frontend techniques they would have otherwise ignored. This reciprocal mentorship model transforms the undergraduate role from a traditional "teaching assistant" to a "technical partner," fostering a more authentic engineering team dynamic.

### *IV.AI Amplifies Mentoring Demands*

Contrary to the common expectation that Large Language Models (LLMs) foster student autonomy, our findings suggest that AI-assisted development actually amplifies the demand for high-level human scaffolding as shown in Figure VIII. While AI tools successfully automated the syntax of the LuckyTag system, they frequently introduced subtle "logical gaps" and security vulnerabilities that the high school novices were unequipped to identify [24]. This accelerated development cycle meant that undergraduate mentors had to provide constant, high-velocity oversight to ensure architectural integrity.

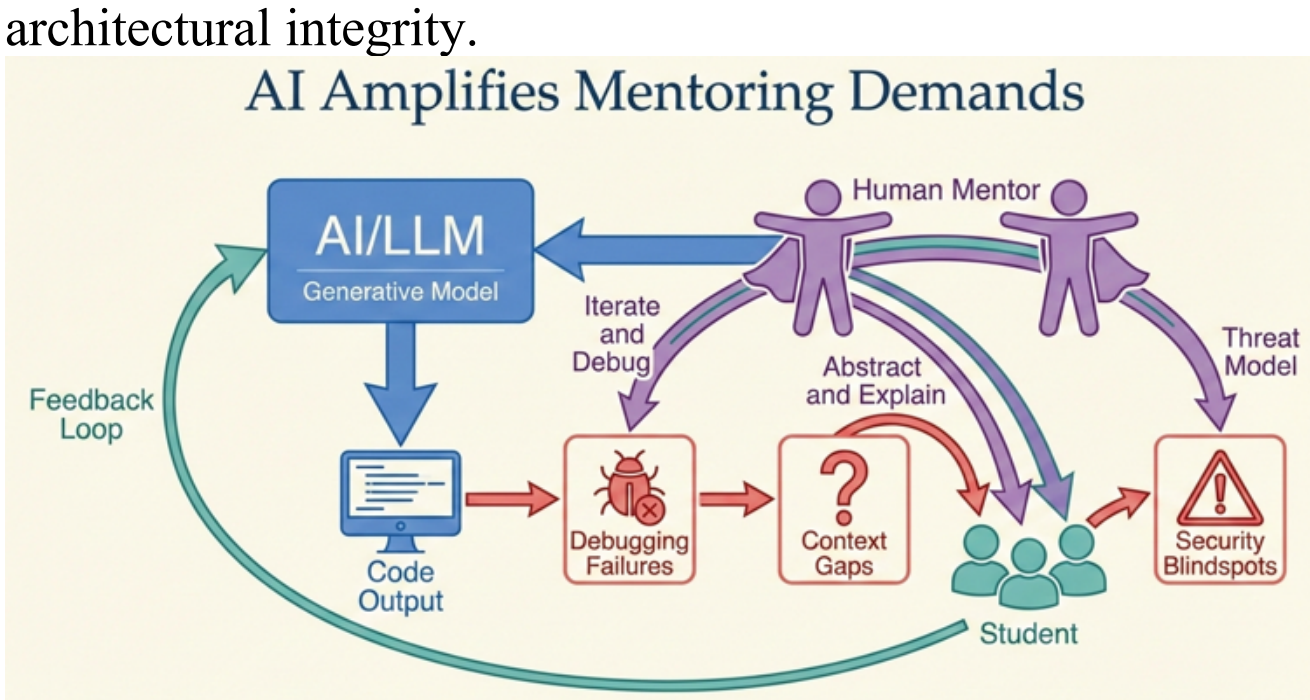

FIGURE VIII
AI AMPLIFIES MENTORING DEMANDS

The QR-to-NFC pivot serves as a definitive example of this dynamic. While the AI tools provided the code for a QR-based system without hesitation, they lacked the Threat Modeling capabilities to recognize the physical security risks (e.g., "QRjacking") and durability concerns inherent in the project's requirements. It was the undergraduate mentors who had to perform the "architectural abstraction", explaining the vulnerability at a conceptual level without jargon, to guide the high school students toward a more robust NFC solution. In this framework, AI does not replace the mentor; instead, it shifts the mentor's burden from "syntax correction" to "security and systems auditing," a task that requires a much higher order of technical leadership.

### *V. Implications for AI-Augmented STEM Education*

Rather than deskilling, AI prompted identity evolution toward oversight roles. One student described engineers as people "managing AI and double-checking" its output. Undergraduates affirmed enduring architecture demands, noting that engineers still need "deep knowledge" to ensure systems remain "scalable and secure." These hybrid roles preserve human agency amid automation.

The failure case in Switzerland highlights the potential for this model to democratize engineering. The high school student possessed the intuition and requirement knowledge but was blocked by the syntax barrier. By utilizing AI-assisted development within a multi-tier mentorship model, we proved that high school students can lead authentic engineering activities that result in deployable, security-conscious systems like LuckyTag. This model allows stem education to focus on the "why" and "what" of engineering, while undergraduate mentors and AI tools handle the "how."

## LIMITATIONS AND FUTURE WORK

The exploratory design precludes causal claims. The small sample (n=9), post-only measurement without baselines, and enthusiastic volunteers introduce selection bias. Self-reports may inflate perceived gains, while one-semester duration limits retention assessment.

Future research will focus on scaling this model to larger, more diverse cohorts to assess its efficacy across different socioeconomic and educational backgrounds. We plan to implement a "Transfer of Learning" assessment to determine if high school students can apply the architectural principles learned through AI-assisted development to traditional, syntax-heavy programming environments. A significant next step is the campus-wide deployment of LuckyTag at Monmouth University. This phase will transition the project from a pilot study to a longitudinal engineering evaluation, providing empirical data on system reliability, finder behavior, and the durability of the NFC hardware. Furthermore, we intend to formalize the "Fundamentals-to-AI" pipeline into a modular curriculum that other universities can adopt to facilitate their own K-12 outreach initiatives.

## CONCLUSION

This study introduces and validates a multi-tiered mentorship framework that integrates AI-assisted development into authentic K-12 engineering activities. By positioning high school students as Product Leads and undergraduate mentors as Technical Architects, we successfully bridged the "syntax barrier" that frequently excludes novices from complex software engineering. The resulting LuckyTag system serves as a proof-of-concept for privacy-preserving, hardware-integrated solutions built through human-AI collaboration.

The findings challenge the prevailing narrative that generative AI "deskills" students. Instead, we observed an identity evolution toward the AI Micromanager, where students focus on logical verification and system-level oversight. Furthermore, the discovery of Reciprocal Mentorship suggests that these collaborations are not merely a service provided to K-12 students, but a valuable leadership laboratory for undergraduate computing majors. As AI continues to lower technical entry barriers, structured mentorship models like the one presented here will be essential to ensure that the next generation of engineers moves beyond passive consumption and toward robust, security-conscious technical leadership.

## References

[1] S. Stavrianeas and M. Stewart. 2022. “Attracting Underrepresented Pre-College Students to STEM Disciplines.” Electronic Journal for Research in Science & Mathematics Education, vol. 26, no. 1, pp. 84–101.

[2] M. J. Hansen, M. J. Palakal, and L. White. Aug. 2024. “The Importance of STEM Sense of Belonging and Academic Hope in Enhancing Persistence for Low-Income, Underrepresented STEM Students.” Journal for STEM Educ Res, vol. 7, no. 2, pp. 155–180, doi: 10.1007/s41979-023-00096-8.

[3] B. Chen, J. Chen, M. Wang, C.-C. Tsai, and P. A. Kirschner. Feb. 2025. “The Effects of Integrated STEM Education on K12 Students’ Achievements: A Meta-Analysis.” Review of Educational Research, p. 00346543251318297, doi: 10.3102/00346543251318297.

[4] National Research Council, et al. 2011. “Successful K-12 STEM Education: Identifying Effective Approaches in Science, Technology, Engineering, and Mathematics.” National Academies Press.

[5] Sowmiya, K., B. Giridharan, and A. Shalini. Mar. 2025.“Artificial Intelligence-Based Student Computational Thinking Analysis Model (AI-SCTAM) for Curriculum Development.” in 2025 International Conference on Machine Learning and Autonomous Systems (ICMLAS), pp. 1292–1299. doi: 10.1109/ICMLAS64557.2025.10968634.

[6] A. Byars-Winston and M. L. Dahlberg. 2019. “The Science of Effective Mentorship in STEMM. Consensus Study Report.” National Academies Press.

[7] J. Prather et al.. Aug. 2024. “The Widening Gap: The Benefits and Harms of Generative AI for Novice Programmers.” in Proceedings of the 2024 ACM Conference on International Computing Education Research - Volume 1, Melbourne VIC Australia: ACM, pp. 469–486. doi: 10.1145/3632620.3671116.

[8] D. Nally. Nov. 2025. “AI-Informed Pedagogy for a Post-Truth Era.” Digit. Soc., vol. 4, no. 3, p. 76, doi: 10.1007/s44206-025-00230-6.

[9] M. R. Morris. Dec. 2024. “Prompting Considered Harmful.” Commun. ACM, vol. 67, no. 12, pp. 28–30, doi: 10.1145/3673861.

[10] M. A. Ferrag et al.. Apr. 2025. “SecureFalcon: Are We There Yet in Automated Software Vulnerability Detection With LLMs?” IEEE Transactions on Software Engineering, vol. 51, no. 4, pp. 1248–1265, doi: 10.1109/TSE.2025.3548168.

[11] M. E. Tuck, K. A. Palomino, J. A. Bradley, et al.. Feb. 2025. “A Coaching-Based Training for Underrepresented Mentors in STEM.” Education Sciences, vol. 15, no. 3, p. 289, doi: 10.3390/educsci15030289.

[12] M. Khalafalla, T. Mulay, D. Regalado, et al.. Jun. 2025. “Decarbonization Education for K-12: A Pilot Study on Transforming Student Perceptions and Career Trajectories in Clean Energy.” in 2025 ASEE Annual Conference & Exposition Proceedings, Montreal, Quebec, Canada: ASEE Conferences, p. 56208. doi: 10.18260/1-2--56208.

[13] E. B. Raposa et al.. Mar. 2019. “The Effects of Youth Mentoring Programs: A Meta-analysis of Outcome Studies.” J Youth Adolescence, vol. 48, no. 3, pp. 423–443, doi: 10.1007/s10964-019-00982-8.

[14] S. Castañeda-Burciaga et al.. Jan. 2026. “Towards an Integrated Educational Practice: Application of Systems Thinking in STEM Disciplines,” Systems, vol. 14, no. 1, p. 97, doi: 10.3390/systems14010097.

[15] S. Sarsa, P. Denny, A. Hellas, and J. Leinonen. Aug. 2022. “Automatic Generation of Programming Exercises and Code Explanations Using Large Language Models,” in Proceedings of the 2022 ACM Conference on International Computing Education Research - Volume 1, Lugano and Virtual Event Switzerland: ACM, pp. 27–43. doi: 10.1145/3501385.3543957.

[16] V. Terragni, A. Vella, P. Roop, and K. Blincoe. Jun. 2025. “The Future of AI-Driven Software Engineering,” ACM Trans. Softw. Eng. Methodol., vol. 34, no. 5, pp. 1–20, doi: 10.1145/3715003.

[17] R. Y. Pang et al.. Apr. 2025. “Understanding the LLM-ification of CHI: Unpacking the Impact of LLMs at CHI through a Systematic Literature Review,” in Proceedings of the 2025 CHI Conference on Human Factors in Computing Systems, Yokohama Japan: ACM, pp. 1–20. doi: 10.1145/3706598.3713726.

[18] N. Perry, M. Srivastava, D. Kumar, and D. Boneh. Nov. 2023. “Do Users Write More Insecure Code with AI Assistants?” in Proceedings of the 2023 ACM SIGSAC Conference on Computer and Communications Security, Copenhagen Denmark: ACM, pp. 2785–2799. doi: 10.1145/3576915.3623157.

[19] X. (Lance) Gu and B. J. Ericson. Aug. 2025. “AI Literacy in K-12 and Higher Education in the Wake of Generative AI: An Integrative Review.” in Proceedings of the 2025 ACM Conference on International Computing Education Research V.1, Charlottesville USA: ACM, Aug. 2025, pp. 125–140. doi: 10.1145/3702652.3744217.

[20] L. Zhang and Y. Ma. Jul. 2023. “A study of the impact of project-based learning on student learning effects: a meta-analysis study.” Front. Psychol., vol. 14, doi: 10.3389/fpsyg.2023.1202728.

[21] H. Cao. Aug. 2025. “Experiences and Insights Gained from AI-Assisted Programming Instruction in Higher Education.” in Proceedings of the 2025 International Conference on AI-enabled Education, Qingdao China: ACM, pp. 159–166. doi: 10.1145/3768421.3768449.

[22] L. H. Silva, R. X. Castro, and M. C. Guimaraes. May 2021. “Supporting Real Demands in Software Engineering with a Four Steps Project-Based Learning Approach.” in 2021 IEEE/ACM 43rd International Conference on Software Engineering: Software Engineering Education and Training (ICSE-SEET), pp. 50–59. doi: 10.1109/ICSE-SEET52601.2021.00014.

[23] P. Robe, S. K. Kuttal, J. AuBuchon, and J. Hart. Nov. 2022. “Pair programming conversations with agents vs. developers: challenges and opportunities for SE community.” in Proceedings of the 30th ACM Joint European Software Engineering Conference and Symposium on the Foundations of Software Engineering, Singapore Singapore: ACM, pp. 319–331. doi: 10.1145/3540250.3549127.

[24] Li, B., Buzaid, C. and Qu, W., 2025. “Security Education in Higher Education through AI-Powered Gamification.” Journal of Cybersecurity, Digital Forensics and Jurisprudence, 1, pp.65-80. doi: 10.65879/3070-5789.2025.01.07